\documentclass{firered}
\usepackage[utf8]{inputenc} 
\usepackage[T1]{fontenc}    
\usepackage{hyperref}       
\usepackage{url}            
\usepackage{booktabs}       
\usepackage{amsfonts}       
\usepackage{nicefrac}       
\usepackage{microtype}      
\usepackage{float}
\usepackage{graphicx}
\usepackage{amssymb}
\usepackage{array}
\usepackage{multirow}
\usepackage{makecell}
\usepackage{caption}
\usepackage{subcaption}
\usepackage{amsmath}
\usepackage{xspace}
\usepackage{tocloft}  
\usepackage{fancyhdr} 
\usepackage[numbers]{natbib}  

\fancypagestyle{firststyle}{
    \fancyhead[L]{}
    \fancyhead[C]{}
    \fancyhead[R]{}
    
    \renewcommand{\footrulewidth}{0.8pt}
    \renewcommand{\headrule}{}
    \renewcommand{\footrule}{\color{sectionred}\hrule width\textwidth height\footrulewidth \vspace{-100pt}}
    \fancyfoot[L]{\color{black}\footnotesize Corresponding author: Kunyang Zhou.
    E-mail: \href{mailto:zhoukunyang1@xiaohongshu.com}{zhoukunyang1@xiaohongshu.com}.}
    \fancyfoot[C]{}
}

\renewcommand{\headrule}{}
\renewcommand{\footrulewidth}{0pt}

\usepackage{tikz}
\usepackage{geometry}
\usepackage{xcolor}

\definecolor{fireredlogo}{RGB}{214,96,87}

\definecolor{scholarblue}{rgb}{0.21,0.49,0.74}
\definecolor{bluelink}{RGB}{0,113,188}
\hypersetup{
    colorlinks=true,%
    citecolor=scholarblue,%
    filecolor=red,%
    linkcolor=red!93!black,%
    urlcolor=bluelink,%
    pdftitle={RedMall: Generative Recommendation for Large-Scale E-commerce},%
    pdfauthor={Trovato et al.}
}

\usepackage{titlesec}
\titlespacing*{\paragraph}
    {0pt}     
    {0.25em}  
    {1em}  

\definecolor{sectionred}{RGB}{220,38,38}  
\titleformat{\section}{\color{sectionred}\large\bfseries}{\color{sectionred}\thesection.}{0.5em}{#1}[]
\titleformat{\subsection}{\color{sectionred}\bfseries}{\color{sectionred}\thesubsection.}{0.5em}{#1}[]
\titleformat{\subsubsection}{\color{sectionred}\bfseries\itshape}{\color{sectionred}\thesubsubsection.}{0.5em}{#1}[]

\title{\center{RedMall: Generative Recommendation for Large-Scale E-commerce}}

\renewcommand\Affilfont{\normalfont\fontsize{11}{15}\selectfont\centering}
\author[1]{Kunyang Zhou}
\author[1]{Xin Jia}
\author[2]{Yi Yang}
\author[1]{Yongjun Yang}
\author[1]{Manjie Xu}
\author[1]{Honglei Zhang}
\author[1]{Yan Zhu}
\author[1]{Kai Zuo}
\author[1]{Xu Tang}
\author[2]{Zheng Wang}

\affil[1]{Xiaohongshu Inc}
\affil[2]{Tongji University}

\begin{document}

{\color{sectionred}\hrule width\textwidth height0.2pt}

\renewcommand{\abstractname}{\textcolor{sectionred}{Abstract}}
\begin{abstract}
Generative recommendation has shown considerable potential in industrial recommendation systems. However, the performance of existing generative retrieval methods is limited by several challenges. \textbf{1)} Semantic IDs (SIDs) are treated as one-dimensional sequences, overlooking the hierarchical dependencies among different layers. 
\textbf{2)} Next-token prediction (NTP) used in existing methods is unable to explicitly optimize session-wise recommendation. 
\textbf{3)} Although reinforcement learning (RL) has been applied to align generated recommendations with user preferences, few existing methods explicitly optimize the ranking of positive items in session-wise recommendation.

To address the above issues, we propose RedMall, a generative recommender designed for large-scale e-commerce. 
Firstly, RedMall constructs hierarchical SIDs through residual quantization over multimodal representations. Secondly, RedMall introduces a lightweight SID encoder to explicitly model layer-wise dependencies in SIDs. 
Moreover, RedMall incorporates sparse mixture-of-experts, dual gated attention, and multi-target prediction (MTP) to scale model size, enhance user interest representation, and optimize session-wise recommendation. 
We further design an exposure-aware loss that treats exposed but unclicked items as negative feedback. To better align user preferences and list-wise ranking objectives, we combine Group Relative Policy Optimization (GRPO) with an auxiliary LambdaRank loss, 
using both real user feedback and item ranking to guide optimization. Experiments on four public datasets---Amazon Beauty, Amazon Toys, Amazon Food, and MovieLens-1M---show that RedMall outperforms the evaluated baselines across all reported Recall and NDCG metrics. 
Online A/B tests show improvements of 2.24\% in advertiser value (ADVV) and 0.59\% in click-through rate (CTR) for e-commerce advertising, 
as well as gains of 1.08\% in GMV and 0.95\% in daily active buyers (DAB) for e-commerce recommendations. These results demonstrate the effectiveness of RedMall for large-scale generative recommendation in e-commerce and advertising.


\end{abstract}

\setlength{\parindent}{0pt}

\begin{tikzpicture}[remember picture,overlay]
  \node[anchor=north west, xshift=2cm, yshift=-1.3cm] at (current page.north west) {%
    \raisebox{-0.28\height}{\includegraphics[height=0.9cm]{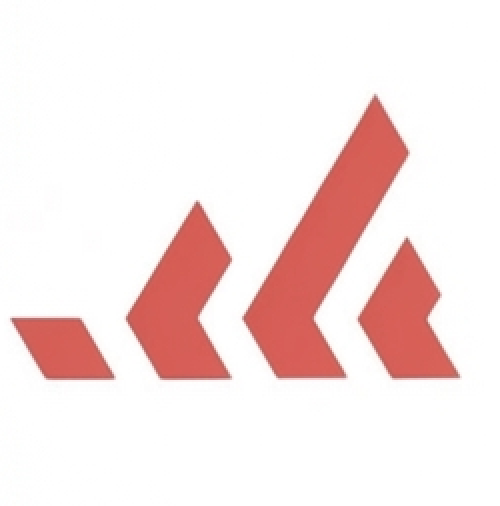}}%
    \hspace{0.15em}%
    {\fontsize{22}{22}\selectfont\bfseries\color{fireredlogo}FireRed}%
    \hspace{0.25em}{\fontsize{22}{22}\selectfont\color{fireredlogo}$\vert$}\hspace{0.25em}%
    {\fontsize{22}{22}\selectfont\bfseries\color{fireredlogo}RedMall}%
  };
\end{tikzpicture}

\begin{figure}[b!] 
  \centering 
  \includegraphics[width=\textwidth]{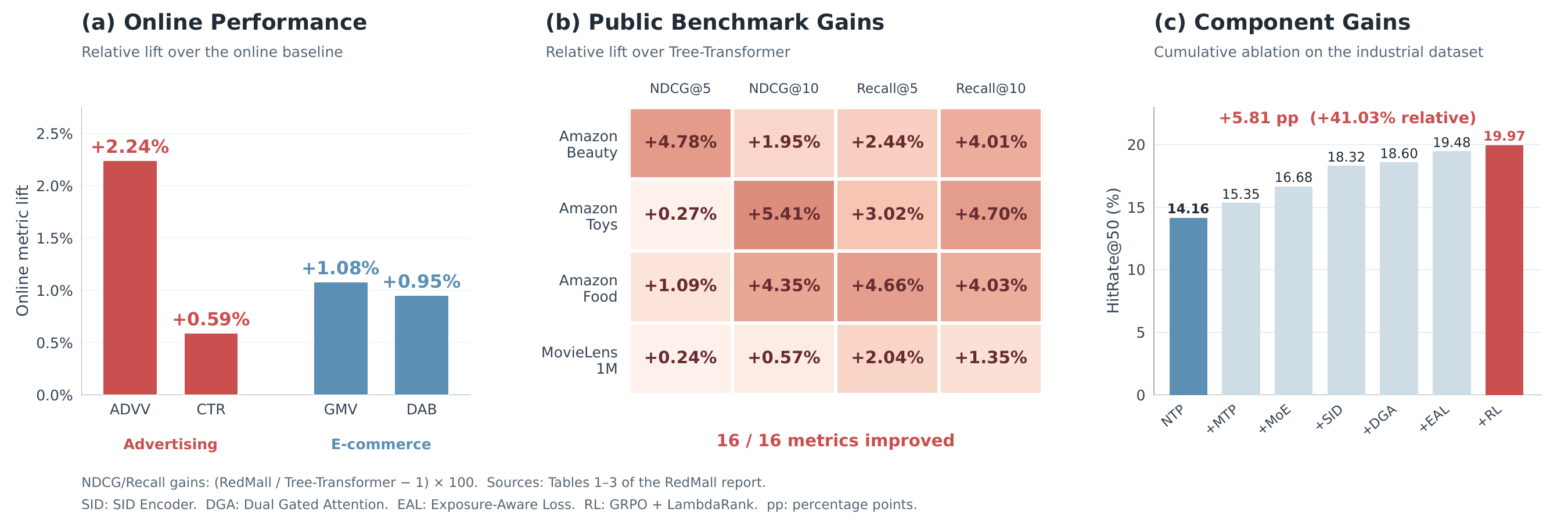} 
  \captionsetup{justification=centering}
  \caption{Online performance, public benchmark results, and component gains.} 
  \label{fig:test2} 
\end{figure}

\maketitle

\setcounter{footnote}{0}
\renewcommand{\thefootnote}{\arabic{footnote}}

\clearpage
\renewcommand{\contentsname}{\textcolor{sectionred}{Contents}}
\tableofcontents
\clearpage

\section{Introduction}

Large-scale e-commerce platforms serve millions of users
and manage vast, continuously evolving item catalogs,
making personalized recommendation essential for
efficient product discovery and user engagement.
Sequential recommendation plays a crucial role in
these platforms by predicting users' future interests
from their chronological interaction histories.
Conventional sequential recommenders, such as
GRU4Rec, SASRec, and BERT4Rec, employ recurrent
or Transformer-based architectures to model
user behavior
\cite{hidasi2015session,kang2018self,sun2019bert4rec}.
Despite their effectiveness, these methods typically
rely on scoring items from a predefined catalog,
which poses challenges for efficient retrieval
as the item space and recommendation scenarios
continue to expand.

Generative recommendation offers an alternative by representing items as sequences of discrete semantic tokens and casting retrieval as conditional sequence generation. TIGER~\cite{rajput2023recommender} established this paradigm by constructing semantic IDs (SIDs) with residual quantization and generating the next item autoregressively. Subsequent work has explored improved item tokenization and recommendation architectures, including incorporating collaborative signals into item representations \cite{wang2024learnable,liu2025endtoend}, as well as modeling the structure of semantic tokens \cite{xu2026trie}. This paradigm provides a unified way to represent items and generate recommendations, and has shown promise for large-scale retrieval.


Despite recent progress, applying generative recommendation to large-scale e-commerce still faces several challenges:

\begin{itemize}
    \item \textbf{Hierarchical dependencies in semantic IDs.}
    A SID is not merely a flat token sequence: its levels form a hierarchy, where later codes are generated from the earlier ones. Treating SIDs as ordinary one-dimensional sequences may prevent the model from explicitly learning these dependencies.

    \item \textbf{Limited supervision for session-wise recommendation.}
    Standard next-token prediction (NTP) supervises a single subsequent item for each user context, whereas practical recommendation often needs to retrieve multiple relevant items. This gap can limit the diversity of training signals and weaken the alignment between the training objective and list-based retrieval.

    \item \textbf{Underuse of exposure-based negative feedback.}
    Click-only supervision overlooks information in exposure logs. Items that are shown to users but not clicked can provide informative negative feedback, yet this signal is not directly captured by standard NTP.

    \item \textbf{Insufficient listwise ranking optimization.}
    Reinforcement learning (RL) can align generated recommendations with user feedback. However, optimizing candidate-level rewards alone does not explicitly encourage a better relative ordering of candidates within a recommendation list.
\end{itemize}

We propose \textbf{RedMall}, a generative recommender designed for large-scale e-commerce that addresses these challenges through coordinated designs in item tokenization, model architecture, and learning objectives. 
RedMall constructs hierarchical SIDs by applying residual vector quantization to multimodal item representations refined with chronological behavior supervision. To expose the structure of these SIDs to the recommender, 
we introduce a lightweight SID encoder that fuses embeddings across SID levels and maps the fused representation back to level-specific features. The generation model further combines a sparse mixture-of-experts (MoE) decoder, 
multi-target prediction (MTP), and dual gated attention. MTP supports session-wise recommendation, while dual gated attention adaptively modulates attention outputs and enables bidirectional modeling of the observed interaction history.

To use behavioral feedback beyond clicks, we introduce an exposure-aware loss that treats exposed but unclicked items as negative feedback during training. We further combine Group Relative Policy Optimization (GRPO) \cite{shao2024deepseekmath} with an auxiliary LambdaRank objective \cite{wang2018lambdaloss}. 
The reward combines real-world user feedback with
ranking model predictions to guide preference alignment,
while the auxiliary LambdaRank loss explicitly optimizes
the relative ordering of generated items.
Together, these objectives improve recommendation
relevance and list-wise ranking quality.

We evaluate RedMall on four public benchmarks---Amazon Beauty, Amazon Toys, Amazon Food, and MovieLens-1M---and conduct online A/B experiments in industrial e-commerce recommendation and advertising scenarios. RedMall outperforms the evaluated baselines across all reported Recall and NDCG metrics on the public datasets. In online experiments, it improves advertiser value by 2.24\% and click-through rate by 0.59\% in e-commerce advertising; in e-commerce recommendation, it improves gross merchandise volume by 1.08\% and daily active buyers by 0.95\%. These results provide evidence for the effectiveness of RedMall in both offline sequential recommendation and real-world large-scale retrieval.

The contributions of this work are summarized as follows:
\begin{itemize}
    \item We present RedMall, a generative recommender for large-scale e-commerce that integrates multimodal, behavior-aware SID construction with generative retrieval.
    \item We introduce a lightweight SID encoder to explicitly model dependencies across SID levels, and combine it with MoE, MTP, and dual gated attention to strengthen representation learning and multi-item recommendation.
    \item We propose an exposure-aware training objective and a preference-alignment framework that combines GRPO with LambdaRank to incorporate negative exposure feedback and improve the relative ordering of generated candidates.
    \item We conduct experiments on four public datasets and online A/B tests in e-commerce and advertising, demonstrating consistent improvements over the evaluated baselines and gains in business metrics.
\end{itemize}

\section{Related Work}

\noindent\textbf{Sequential Recommendation.}
Sequential recommendation aims to predict users' future interactions based on their historical behavior sequences.
Early neural approaches employ various architectures to capture temporal patterns and sequential dependencies, including GRU4Rec~\cite{hidasi2015session}, SASRec~\cite{kang2018self}, and BERT4Rec~\cite{sun2019bert4rec}.
Recently, advances in large language models (LLMs) have inspired two major research directions: representation-oriented and generation-oriented approaches.
The former leverages the powerful representation capabilities of LLMs to enhance item embeddings~\cite{liu2025llmemb,liu2024llm}, or employs LLMs to model user behavior sequences and predict future item representations~\cite{chen2024hllm,zhou2026hymirec}.
In contrast, generation-oriented approaches reformulate recommendation as an autoregressive generation task, directly generating item identifiers or semantic tokens conditioned on user interaction histories.
This paradigm reduces the reliance on conventional candidate scoring and has motivated recent advances in generative recommendation.

\noindent\textbf{Generative Recommendation.}
Unlike traditional recommendation frameworks, generative recommendation represents items as discrete token sequences and formulates recommendation as a sequence generation problem.
Existing methods typically consist of two key components: item tokenization and generative modeling.
TIGER~\cite{rajput2023recommender} pioneered Semantic ID (SID)-based generative recommendation by employing RQ-VAE~\cite{lee2022autoregressive} for item tokenization and T5~\cite{raffel2020exploring} for autoregressive item generation.
Subsequent studies have improved item representations and tokenization strategies.
LETTER~\cite{wang2024learnable} and CoST~\cite{zhu2024cost} incorporate collaborative signals and contrastive learning to enhance semantic representations, while CoFiRec introduces item-aware positional encoding to improve feature modeling.
Recent industrial advances further demonstrate the scalability and effectiveness of generative recommendation.
OneRec~\cite{zhou2025onerec} establishes an end-to-end generative recommendation framework for large-scale industrial applications, unifying retrieval and ranking through session-wise generation and preference alignment.
OneRec-V2~\cite{Zhou2025OneRecV2TR} improves computational efficiency through a lazy decoder-only architecture and incorporates reinforcement learning with real-world user feedback for preference alignment.
OneRec-Think~\cite{Liu2025OneRecThinkIR} further explores explicit reasoning by integrating semantic alignment, reasoning, and personalized recommendation.

Despite these advances, most SID-based generative recommenders
treat SIDs as flattened token sequences, overlooking their
inherent hierarchical dependencies.
Tree-Transformer~\cite{xu2026trie} addresses this limitation
by introducing tree-aware absolute positional encoding and
topology-aware relative positional encoding to capture
SID structural information.
However, these specialized positional encodings may compromise
the original sequential positional information.
To overcome this limitation, we propose a lightweight SID encoder
that explicitly captures hierarchical dependencies within SIDs
while preserving the original sequential positional encoding.

\section{Methods}

\subsection{Semantic ID Tokenization}
\label{sec:sid}

Recent generative recommenders represent items as semantic token sequences and formulate retrieval as autoregressive decoding~\cite{rajput2023recommender,wang2024learnable,liu2025endtoend}. Following this paradigm, we construct a hierarchical semantic ID (SID) for each item by coupling sequence-aware multimodal representation learning with residual vector quantization. We adapt the tokenizer to an industrial item corpus: the representation incorporates behavioral transitions while retaining textual and visual evidence, and the quantizer is trained and maintained online as a unified module.

\paragraph{Sequence-aware multimodal representation.} Content encoders provide a text embedding $\mathbf{e}^{t}$ and a visual embedding $\mathbf{e}^{v}$ for each item. Content similarity alone, however, does not necessarily reflect substitutability or transition patterns in user behavior. Recent tokenizers therefore augment content semantics with collaborative supervision~\cite{wang2024learnable,liu2025endtoend}. We post-train the multimodal representations by next-item prediction over chronological interaction sequences. This objective transfers behavioral co-occurrence and transition structure into the item embedding while preserving content features for newly published items.

The two modalities are projected into a common space and fused with a learned gate. This design provides an item-dependent alternative to fixed multimodal fusion, which is particularly relevant when constructing SIDs from heterogeneous content signals~\cite{wang2025mmesid}:
\begin{align}
  \mathbf{h}^{t} &= \phi_t(\mathbf{e}^{t}), \\
  \mathbf{h}^{v} &= \phi_v(\mathbf{e}^{v}), \\
  \mathbf{g} &= \sigma\!\left(
    \mathbf{W}_g[\mathbf{h}^{t};\mathbf{h}^{v}]+\mathbf{b}_g\right), \\
  \mathbf{x} &= \mathbf{g}\odot\mathbf{h}^{t}
    +(\mathbf{1}-\mathbf{g})\odot\mathbf{h}^{v},
  \label{eq:gated_fusion}
\end{align}
where $\phi_t$ and $\phi_v$ are modality-specific projections and $\odot$ denotes element-wise multiplication. Unlike fixed fusion, the gate adapts the contribution of each modality at the item level. The fused vector $\mathbf{x}$ is used as the input and reconstruction target of the SID tokenizer.

\paragraph{End-to-end residual quantization.} We adopt RQ-VAE~\cite{lee2022autoregressive}, which has become a standard backbone for semantic item tokenization~\cite{rajput2023recommender,wang2024learnable}, to map $\mathbf{x}$ to a sequence of discrete codes. Let $\mathbf{r}_1$ be the encoder output. At level $\ell$, the quantizer selects the nearest codeword from codebook $\mathcal{C}_{\ell}$ and passes the residual to the next level:
\begin{equation}
  \mathbf{q}_{\ell}
  = \arg\min_{\mathbf{e}\in\mathcal{C}_{\ell}}
    \|\mathbf{r}_{\ell}-\mathbf{e}\|_2^2,
  \qquad
  \mathbf{r}_{\ell+1}=\mathbf{r}_{\ell}-\mathbf{q}_{\ell}.
  \label{eq:rq}
\end{equation}
The selected codeword indices constitute the SID, with earlier levels capturing coarse semantics and later levels encoding residual variation. We use three levels with 8,192, 1,024, and 128 candidate codes, respectively.

The encoder, quantizer, and decoder are optimized jointly:
\begin{equation}
  \mathcal{L}_{\mathrm{SID}}
  = \|\mathbf{x}-\hat{\mathbf{x}}\|_2^2
  + \lambda_c\sum_{\ell=1}^{L}
    \|\mathbf{r}_{\ell}-
      \operatorname{sg}[\mathbf{q}_{\ell}]\|_2^2,
  \label{eq:sid_objective}
\end{equation}
where $\operatorname{sg}[\cdot]$ denotes stop-gradient. We use the straight-through estimator~\cite{oord2017neural} to propagate reconstruction gradients through discrete assignments. An example of an item SID is shown below:

\begin{equation}
\texttt{\detokenize{<s_a_8000><s_b_1000><s_c_76><item_end>}}
\end{equation}

\paragraph{Codebook stabilization.} Codewords are updated online by exponential moving averages with decay $0.995$. To prevent persistently inactive codes from reducing the effective token space, we re-initialize a codeword from the current mini-batch when its usage ratio falls below $0.2$.

\subsection{Generative Recommendation}

\begin{figure}[t!] 
  \centering 
  \includegraphics[width=0.8\textwidth]{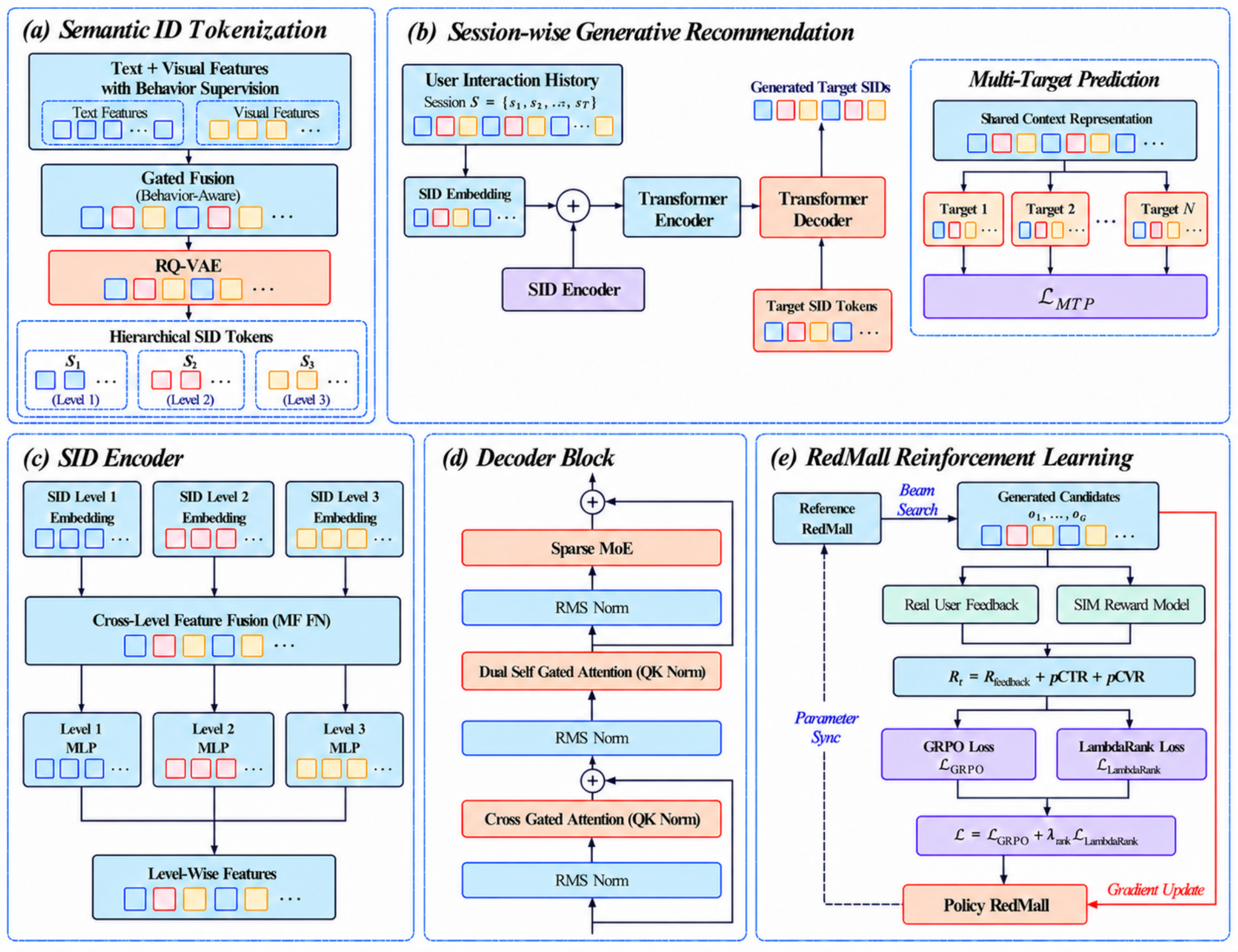} 
  \caption{Overview of the RedMall architecture.
A lightweight SID encoder captures hierarchical dependencies
within semantic IDs and enhances user interaction representations.
The Transformer encoder-decoder architecture integrates
sparse MoE, dual gated attention, and multi-target prediction
for effective generative recommendation.
An exposure-aware loss incorporates negative feedback
from unclicked exposures, while GRPO with an auxiliary
LambdaRank loss aligns user preferences and optimizes
list-wise ranking.} 
  \label{fig:test1} 
\end{figure}

The overall architecture of RedMall is illustrated
in Figure~\ref{fig:test1}.
We first describe the Transformer encoder and decoder
architectures in Sections 3.2.1 and 3.2.3, respectively.
Section 3.2.2 introduces a lightweight SID encoder
that explicitly models hierarchical dependencies
across SID levels.
Finally, Section 3.2.4 presents the reinforcement
learning framework for user preference alignment
and list-wise ranking optimization.

\subsubsection{Encoder Architecture}
\label{sec1}
As illustrated in Figure~\ref{fig:test1}, the encoder processes the user interaction history and produces a user behavior representation using a Transformer-based framework. Each encoder block consists of a fully visible self-attention module and a feed-forward network (FFN), each preceded by RMS normalization. 
First, the user interaction sequence $M$ is passed through the SID embedding layer to obtain the history embeddings, denoted by $\mathbf{a}^{(i)}$. These embeddings are then processed by $L_{N}$ encoder blocks as follows:

\begin{align}
\mathbf{a}^{(i)} &= \mathrm{Embedding}\big(M \big), \\
\mathbf{a}^{(i+1)} &= \mathbf{a}^{(i)} + \mathrm{SelfAttn}\big(\mathrm{RMSNorm}(\mathbf{a}^{(i)})\big), \\
\mathbf{a}^{(i+1)} &= \mathbf{a}^{(i+1)} + \mathrm{FFN}\big(\mathrm{RMSNorm}(\mathbf{a}^{(i+1)})\big).
\end{align}

The output of the final encoder block $\mathbf{a}^{L_{N}}$ denotes the user behavior representation, serving as the basis for generation.

\subsubsection{SID Encoder}
\label{sec4}
\textbf{Motivation}. Existing generative recommendation methods such as OneRec~\cite{zhou2025onerec} and OneMall~\cite{zhang2026onemall} treat the input as a 1D sequence, ignoring that
SIDs have a hierarchical dependency structure: the second layer depends on the first, and the third layer depends on the preceding layers. 
However, next-token prediction alone does not explicitly model these hierarchical dependencies.

We propose a lightweight SID encoder, a plug-and-play module, to learn layer-wise dependencies within SIDs. As shown in Figure~\ref{fig:test1} (c), we first obtain the embedding of each SID layer, 
using an embedding layer that shares weights with the encoder's SID embedding layer. Then, we adopt a modified feed-forward network (MFFN) to achieve cross-layer feature fusion along the layer dimension. Finally, we use three MLP layers 
to map the fused SID embedding to layer-specific representations. The process of the SID encoder can be written as:

\begin{align}
\mathbf{s}^{1}, \mathbf{s}^{2}, \mathbf{s}^{3} &= \mathrm{Embedding}(\mathrm{SID}), \\
\mathbf{s}^{\mathrm{fuse}} &= \mathrm{MFFN}\big(\mathbf{s}^{1}, \mathbf{s}^{2}, \mathbf{s}^{3}\big), \\
\mathbf{h}^{(i)} &= \mathrm{MLP}_{i}\big(\mathbf{s}^{\mathrm{fuse}}\big), \quad i = 1, 2, 3.
\end{align}

where $\mathbf{s}^{1}, \mathbf{s}^{2}, \mathbf{s}^{3}$ are the embeddings of the three SID layers. The MFFN consists of layer normalization, an MLP layer that increases the dimensionality, a GELU activation, an MLP layer that reduces the dimensionality, and a final layer normalization. $\mathbf{h}^{(i)}$ is added to $\mathbf{a}^{(i)}$, which is taken as the input of the encoder.

\subsubsection{Decoder Architecture}
\label{sec2}
The architecture of a single decoder block is depicted in Figure~\ref{fig:test1} (d). The inputs to the decoder consist of two components: the hidden state from the encoder and the target item. We adopt the decoder in TIGER~\cite{rajput2023recommender}
as the base architecture. However, compared with advanced decoders in modern LLMs such as Qwen3~\cite{yang2025qwen3}, TIGER's vanilla decoder has limited capacity. We introduce several modules to enhance the decoder.

\textbf{Mixture of Experts}. The small size of TIGER's decoder constrains its capacity to learn expressive feature representations. Inspired by the architectural design of large language models, we introduce a sparse mixture-of-experts (MoE) architecture to increase model capacity. 
Specifically, we replace the feed-forward sublayers in the decoder with MoE modules, each comprising 12 experts, with only two experts activated during inference. The resulting model contains approximately 0.5 billion total parameters and 0.1 billion active parameters, denoted as 0.5B-A0.1B.

Each expert is a gated feed-forward network with pre-normalization
and a residual connection. Given an input tensor
$\mathbf{X} \in \mathbb{R}^{N \times d_{\mathrm{model}}}$,
we first apply RMSNorm, followed by two parallel, bias-free linear
projections into an intermediate space of dimension $d_{\mathrm{ff}}$.
One branch applies a nonlinear activation, and its output is
multiplied element-wise with the output of the other branch.
The resulting representation passes through dropout, an output
projection, and a second dropout operation before being added
to the original input:
\begin{align}
\mathbf{Z}_e
&= \operatorname{RMSNorm}_e(\mathbf{X}), \\
\mathbf{A}_e
&= \phi\!\left(\operatorname{Linear}_{e,1}(\mathbf{Z}_e)\right), \\
\mathbf{B}_e
&= \operatorname{Linear}_{e,2}(\mathbf{Z}_e), \\
\mathbf{H}_e
&= \mathbf{A}_e \odot \mathbf{B}_e, \\
\mathbf{U}_e
&= \operatorname{Linear}_{e,o}\!\left(
    \operatorname{Dropout}(\mathbf{H}_e)
\right), \\
E_e(\mathbf{X})
&= \mathbf{X} + \operatorname{Dropout}(\mathbf{U}_e).
\end{align}
Here, $e$ indexes the expert, $\phi$ denotes the activation function,
and $\odot$ denotes element-wise multiplication.
Both $\operatorname{Linear}_{e,1}$ and
$\operatorname{Linear}_{e,2}$ map $d_{\mathrm{model}}$ to
$d_{\mathrm{ff}}$, while $\operatorname{Linear}_{e,o}$ maps
$d_{\mathrm{ff}}$ back to $d_{\mathrm{model}}$.
The MoE loss is the same as that used in Qwen3.

\textbf{Multi-Target Prediction}. Generative retrieval methods, such as TIGER and OneRec,
formulate recommendation as autoregressive next-token prediction
(NTP). However, in modern recommendation scenarios,
a user's recent interaction history can support multiple relevant
recommendations, whereas supervision based on a single subsequent
item provides a limited training signal. We therefore introduce
multi-target prediction (MTP), extending single-item supervision
to multiple target items under the same user context.
Compared with single-target NTP training, MTP provides richer
supervision that may capture a broader range of user interests
and reduce reliance on a single observed interaction.
It also better aligns training with session-wise recommendation,
where multiple relevant items must be retrieved for each
user context.

Specifically, we reserve the final $N$ items of each training
sequence as prediction targets and use the preceding interactions
as the shared user context. The decoder is trained separately
for each target item, resulting in $N$ forward passes per training
example. Each pass retains the autoregressive token-level
prediction objective for its corresponding item, and the losses
are summed across all $N$ targets:
\begin{equation}
\mathcal{L}_{\mathrm{MTP}}
=
-\sum_{n=1}^{N}
\sum_{j=1}^{L_n}
\log p_{\theta}\!\left(
y_{n,j}
\mid
\mathbf{hs}, y_{n,<j}
\right),
\end{equation}
where $\mathbf{hs}$ denotes the shared interaction history,
$y_{n,j}$ is the $j$-th token in the representation of the
$n$-th target item, and $L_n$ is its token sequence length.
We set $N=5$ in our experiments.

\textbf{Dual Gated Attention}. We incorporate gated attention~\cite{yang2025qwen3} into the decoder to adaptively
modulate attention outputs and enhance its capacity to learn
informative representations. We further replace causal masking
with bidirectional attention over observed historical items.
Unlike autoregressive token generation, where future target
tokens must remain inaccessible, recommendation conditions
on a fully observed interaction history, allowing historical
items to attend to one another without violating the prediction
boundary. Bidirectional attention therefore enables each
historical item to be contextualized by both earlier and later
interactions, facilitating the modeling of dependencies across
sequence positions and potentially yielding more coherent
representations of user interests. Consistent with this
motivation, we empirically observe improved sequence modeling
with bidirectional history attention (see Table~\ref{tab:ablation-industrial}).

\textbf{Exposure-Aware Loss}. Existing generative recommendation methods typically
rely on next-token prediction (NTP) to optimize the
likelihood of users' interacted items.
However, this objective primarily focuses on positive
interactions while overlooking valuable negative feedback
from exposed but unclicked items.
Such feedback provides additional signals for distinguishing
users' preferred items from less relevant alternatives.
To leverage this information, we introduce an exposure-aware
loss that explicitly incorporates exposed but unclicked
items as negative feedback, enhancing the model's ability
to discriminate between preferred and non-preferred items.

For each target interaction, we collect exposed but unclicked
items within a window of size 5 and select their most frequent
first-level SID as the negative target. Let $\mathcal{B}^{+}$
and $\mathcal{B}^{-}$ denote the positive and negative subsets
of a mini-batch, respectively. We define the positive generation
loss and the negative suppression loss as
\begin{align}
\mathcal{L}_{\mathrm{pos}}
&=
-\sum_{i\in\mathcal{B}^{+}}
\sum_{\ell=1}^{3}
\log p_{\theta}\!\left(
s_i^{(\ell)}
\mid \mathcal{H}_i, s_i^{(<\ell)}
\right),
\\
\mathcal{L}_{\mathrm{neg}}
&=
-\sum_{i\in\mathcal{B}^{-}}
\log\!\left[
1-p_{\theta}\!\left(
\bar{s}_i^{(1)} \mid \mathcal{H}_i
\right)
\right].
\end{align}
The overall exposure-aware objective is
\begin{equation}
\mathcal{L}_{\mathrm{EA}}
=
\frac{
\mathcal{L}_{\mathrm{pos}}
+
\alpha\mathcal{L}_{\mathrm{neg}}
}{
|\mathcal{B}^{+}|+|\mathcal{B}^{-}|
}.
\end{equation}
Here, $\mathcal{H}_i$ denotes the user interaction history,
$s_i^{(\ell)}$ is the positive item's SID token at level $\ell$,
and $\bar{s}_i^{(1)}$ denotes the selected negative SID prefix.
The probability $p_{\theta}$ is obtained from the model's
softmax output, and $\alpha$ controls the strength of negative
supervision. The positive loss encourages generation of the
complete three-level SID, while the negative loss discourages
generation of the selected first-level prefix.

\subsubsection{RedMall Reinforcement Learning}
\label{sec3}
Inspired by user preference alignment in OneRec and list-wise ranking optimization in GR4AD~\cite{xue2026generative}, we introduce reinforcement learning (RL) into the post-training stage to align the model with user preferences through real-world feedback. 
However, user feedback is sparse and delayed, while point-wise rewards alone may not adequately capture the relative quality of candidates within a list. We therefore augment the RL objective with a LambdaRank ranking loss~\cite{wang2018lambdaloss}.
RL improves the policy based on reward-driven group-relative advantages, while the ranking loss provides explicit supervision over the relative ordering of candidates generated for the same request. In this paper, we adopt GRPO~\cite{shao2024deepseekmath} as the RL algorithm.

\textbf{Reward Model}. The online ranking model relies on numerous features, making it inconvenient to use directly for training the generative model. To simplify reward computation, we use the ranking model SIM~\cite{pi2020search} as the reward model. 
Given a user's SID sequence as input, SIM predicts pCTR and pCVR using supervision from the online ranking model.

\textbf{Real-world feedback reward}. Since the reward model may not fully capture users' actual preferences, we also incorporate online user logs. Specifically, clicked samples observed within a future window (we set the window size to 5 in our experiments) are treated as positive samples. Let the predicted SID sequence be
$\hat{y}=(\hat{s}_1,\hat{s}_2,\hat{s}_3)$, and let $\mathcal{Y}^{+}$ denote the set of clicked SID sequences within the window. We define the feedback reward as
\begin{equation}
R_{\mathrm{feedback}}(\hat{y})
=
\begin{cases}
\displaystyle
\max_{y^{+}\in\mathcal{Y}^{+}}
\sum_{t=1}^{3}
\mathbb{I}\!\left[\hat{s}_t=s^{+}_t\right],
& \mathcal{Y}^{+}\neq\emptyset,\\[8pt]
0, & \mathcal{Y}^{+}=\emptyset,
\end{cases}
\label{eq:feedback-reward}
\end{equation}
where $y^{+}=(s^{+}_1,s^{+}_2,s^{+}_3)$ and $\mathbb{I}[\cdot]$ is the indicator function. Each SID level is matched independently, and each matching level contributes one reward point. Thus, the feedback reward of a SID ranges from 0 to 3.

\textbf{Overall Reward}. 
For candidate $\hat{y}_i$, we combine the feedback reward with the reward model predictions:
\begin{equation}
R_i
=
R_{\mathrm{feedback}}(\hat{y}_i)
+
\alpha\,\mathrm{pCTR}(\hat{y}_i)
+
\beta\,\mathrm{pCVR}(\hat{y}_i),
\label{eq:overall-reward}
\end{equation}
where $\alpha$ and $\beta$ control the relative contributions of pCTR and pCVR. We set $\alpha$ and $\beta$ to 1 in our experiments.

\textbf{GRPO with a LambdaRank auxiliary objective}. GRPO uses the overall rewards in Eq.~\ref{eq:overall-reward} to compute group-relative advantages and update the policy. In addition, we apply a LambdaRank loss to the $n$ candidates sampled for the same request. Let
$s_i=\log p_{\theta}(\hat{y}_i\mid x)$ denote the model score of candidate $\hat{y}_i$. For each pair satisfying $R_i>R_j$, the loss encourages the higher-reward candidate to receive a higher model score:
\begin{equation}
\mathcal{L}_{\mathrm{LambdaRank}}
=
\sum_{i=1}^{n}
\sum_{\substack{j=1\\R_i>R_j}}^{n}
\left|\Delta\mathrm{NDCG}_{ij}\right|
\log\!\left(
1+\exp\!\left[-(s_i-s_j)\right]
\right),
\label{eq:lambdarank-loss}
\end{equation}
where $\left|\Delta\mathrm{NDCG}_{ij}\right|$ measures the change in NDCG caused by swapping candidates $i$ and $j$, and serves as the pairwise weight. This weighting assigns greater emphasis to candidate pairs whose ordering has a larger effect on the list-level ranking metric. 
The final training objective is
\begin{equation}
\mathcal{L}
=
\mathcal{L}_{\mathrm{GRPO}}
+
\lambda_{\mathrm{rank}}\mathcal{L}_{\mathrm{LambdaRank}},
\label{eq:combined-objective}
\end{equation}
where $\lambda_{\mathrm{rank}}$ controls the contribution of the LambdaRank auxiliary loss.

\subsection{Online Deployment}

In deploying RedMall to large-scale e-commerce systems,
we address two practical challenges: intra-SID item ranking
and inference acceleration.

\noindent\textbf{Intra-SID Item Ranking.}
Since multiple items may share the same semantic ID (SID),
directly retrieving all items associated with each generated
SID can reduce recommendation diversity and exhaust the
limited retrieval quota, preventing sufficient coverage
of distinct SIDs.
To address this issue, we introduce a posterior-based
ranking strategy to select representative items within
each SID.
Specifically, we assign each item a ranking score
based on two historical performance metrics:
Gross Merchandise Volume per Mille (GPM) and
Orders per Mille (OPM):

\begin{equation}
    score = 0.8 \ln(1+\mathrm{GPM})
        + 0.2 \ln(1+\mathrm{OPM}),
\end{equation}

where per mille means per 1,000 impressions. Items sharing the same SID are ranked by their scores,
and only the top-$K$ items are retained, with $K=40$
in our production deployment.
This strategy improves retrieval quota utilization
and enables broader coverage of distinct SIDs.

\noindent\textbf{Inference Acceleration.}
To support large-scale online serving, we employ
several inference optimizations.
First, we introduce KV caching to accelerate
autoregressive decoding during beam search by
reusing previously computed attention states.
Furthermore, we adopt batched inference and
leverage multithreading and multiprocessing
to accelerate data processing and model inference.
Our production serving system is deployed on
24 NVIDIA L20 GPUs and processes approximately
120 million user inference requests per day.
With these optimizations, the serving system
achieves a throughput of up to 420 queries per
second (QPS) per GPU.

\section{Experiments}

In this section, we conduct comprehensive experiments on public datasets and the industrial dataset to evaluate the effectiveness of our methods. We also conduct online A/B tests to evaluate performance in a real-world recommendation system.

\subsection{Experimental Settings}
\subsubsection{Datasets} We conduct offline experiments on four widely used benchmarks: Amazon-Beauty~\cite{mcauley2013amateurs}, Amazon-Toys~\cite{mcauley2013amateurs}, Amazon-Food~\cite{mcauley2013amateurs}, and MovieLens-1M~\cite{harper2015movielens}. The first three datasets are collected from the Amazon e-commerce platform, containing user ratings on products. MovieLens-1M is a movie rating dataset collected from the MovieLens platform. 
Following previous work~\cite{xu2026trie,rajput2023recommender}, we adopt the standard 5-core preprocessing protocol, retaining the 20 most recent interactions for each user. For evaluation, we employ a leave-one-out partitioning scheme: each user's latest interaction is assigned to the test set, the penultimate interaction to the validation set, and the remaining interactions to the training set. 

\subsubsection{Evaluation Metrics} We evaluate sequential recommendation performance using two widely adopted metrics, Recall@K and NDCG@K, where \(K \in \{5, 10\}\). For conventional recommendation models, scores are computed over the entire item set, and the top-\(K\) ranked items are selected for evaluation. For generative recommendation models, we adopt constrained beam search to guarantee that the generated token sequences can be mapped to valid items. 

\subsubsection{Implementation Details} The encoder and decoder each contain four blocks. The group size of GRPO is 32 and $\lambda_{\mathrm{rank}}$ is set to 0.5. We use Flash Attention 2.0~\cite{dao2022flashattention} to accelerate training and inference. The learning rate is 0.001, and we use the Adam optimizer with a cosine learning rate scheduler. We adopt GRID~\cite{ju2025generative} as our codebase. The beam width is set to 50 for online A/B tests and 10 for public datasets. All experiments are conducted on 56 H800 GPUs. 
For all conventional sequential recommendation methods, we follow the implementation details and hyperparameter settings specified in their original publications. To ensure a fair comparison, we exhaustively search the hyperparameter ranges recommended by the respective authors and, where necessary, further extend these ranges. The best-performing configuration on the validation set is selected for each baseline method.

\subsubsection{Baselines} In order to validate how well our method performs, we compare it with a set of competitive baselines for sequential recommendation, covering widely‑adopted ID‑based and SID-based approaches, i.e., SASRec~\cite{kang2018self}, BERT4Rec~\cite{sun2019bert4rec}, GRU4Rec~\cite{hidasi2015session}, HSTU~\cite{zhai2024actions}, HyMiRec~\cite{zhou2026hymirec}, TIGER~\cite{rajput2023recommender}, and Tree-Transformer~\cite{xu2026trie}.
To achieve a fair comparison, we restrict the input sequence length for HyMiRec to 10 on public benchmarks to limit inference latency following the original paper. Non-LLM baselines are given extended sequences containing 20 items.

\subsection{Offline Experiments}

\begin{table*}[t!]
\centering
\caption{Comparison of recommendation performance across datasets. Boldface marks the best result in each column; N and R denote NDCG and Recall, respectively.}
\label{tab:overall-performance}
\setlength{\tabcolsep}{4pt}
\small
\resizebox{\textwidth}{!}{%
\begin{tabular}{l|cccc|cccc|cccc|cccc}
\toprule
\textbf{Model}
& \multicolumn{4}{c|}{\textbf{Amazon Beauty}}
& \multicolumn{4}{c|}{\textbf{Amazon Toys}}
& \multicolumn{4}{c|}{\textbf{Amazon Food}}
& \multicolumn{4}{c}{\textbf{MovieLens-1M}} \\
\cmidrule(lr){2-5}
\cmidrule(lr){6-9}
\cmidrule(lr){10-13}
\cmidrule(lr){14-17}
& N@5 & N@10 & R@5 & R@10
& N@5 & N@10 & R@5 & R@10
& N@5 & N@10 & R@5 & R@10
& N@5 & N@10 & R@5 & R@10 \\
\midrule
GRU4Rec
& 0.0055 & 0.0079 & 0.0088 & 0.0163
& 0.0045 & 0.0066 & 0.0066 & 0.0131
& 0.0111 & 0.0154 & 0.0178 & 0.0312
& 0.0128 & 0.0199 & 0.0202 & 0.0417 \\
SASRec
& 0.0177 & 0.0263 & 0.0346 & 0.0612
& 0.0229 & 0.0320 & 0.0458 & 0.0739
& 0.0210 & 0.0309 & 0.0411 & 0.0716
& 0.0440 & 0.0641 & 0.0765 & 0.1391 \\
BERT4Rec
& 0.0196 & 0.0269 & 0.0317 & 0.0545
& 0.0207 & 0.0265 & 0.0326 & 0.0507
& 0.0245 & 0.0334 & 0.0399 & 0.0678
& 0.0420 & 0.0597 & 0.0679 & 0.1232 \\
HSTU
& 0.0176 & 0.0264 & 0.0347 & 0.0622
& 0.0238 & 0.0328 & 0.0476 & 0.0754
& 0.0231 & 0.0310 & 0.0416 & 0.0662
& 0.0511 & 0.0719 & 0.0902 & 0.1553 \\
TIGER
& 0.0318 & 0.0395 & 0.0478 & 0.0719
& 0.0332 & 0.0418 & 0.0501 & 0.0765
& 0.0345 & 0.0430 & 0.0522 & 0.0806
& 0.0713 & 0.0912 & 0.1096 & 0.1713 \\
HyMiRec
& 0.0281 & 0.0372 & 0.0410 & 0.0679
& 0.0344 & 0.0426 & 0.0512 & 0.0779
& 0.0337 & 0.0424 & 0.0519 & 0.0792
& 0.0741 & 0.0925 & 0.1103 & 0.1730 \\
Tree-Transformer
& 0.0335 & 0.0410 & 0.0491 & 0.0723
& 0.0368 & 0.0444 & {0.0530} & {0.0766}
& {0.0368} & {0.0460} & {0.0558} & {0.0843}
& {0.0841} & {0.1053} & {0.1272} & {0.1929} \\
\textbf{RedMall}
& \bf 0.0351 & \bf 0.0418 & \bf 0.0503 & \bf 0.0752
& \bf 0.0369 & \bf 0.0468 & \bf 0.0546 & \bf 0.0802
& \bf 0.0372 & \bf 0.0480 & \bf 0.0584 & \bf 0.0877
& \bf 0.0843 & \bf 0.1059 & \bf 0.1298 & \bf 0.1955 \\
\bottomrule
\end{tabular}
}
\end{table*}

\textbf{Results on public benchmarks}. 
Table~\ref{tab:overall-performance} compares RedMall with representative sequential and generative recommendation baselines across four datasets. RedMall achieves the best results on all reported metrics, demonstrating consistent improvements across both NDCG and Recall.
RedMall outperforms Tree-Transformer on all 16 dataset--metric combinations. The gains are consistent across datasets, with relative improvements ranging from 0.14\% to 8.16\%. On Amazon Beauty, RedMall improves N@5 from 0.0335 to 0.0351 and R@10 from 0.0723 to 0.0752.

Compared with TIGER, RedMall also achieves higher scores on every reported metric. The largest improvements are observed on Amazon Food, where N@5 increases from 0.0345 to 0.0372 and R@10 from 0.0806 to 0.0877. RedMall also yields consistent gains on MovieLens-1M, improving N@10 from 0.0912 to 0.1059 and R@10 from 0.1713 to 0.1955. This comparison indicates that RedMall provides stronger recommendation performance than TIGER across both product and movie domains.

\subsection{Online A/B Experiments}
We perform online A/B experiments on downstream recommendation use cases to further demonstrate the real-world effectiveness of RedMall, with all experiments deployed at the recall stage of an industrial recommendation system. The results are shown in Table~\ref{tab:redmall-online-metrics}.

\textbf{E-commerce Advertising}. This scenario centers on standard e‑commerce advertisement recall for commodity feeds. Its primary objective is to improve ADVV (advertiser value). The online A/B test adopts 10\% versus 10\% traffic allocation and lasts for a minimum of seven days. 
Only items posted within the last 24 hours are included in the recall candidate set. RedMall increases ADVV by \textbf{2.24\%} and cost by \textbf{0.56\%}, demonstrating its ability to improve advertising value.

\textbf{E-commerce Recommendation}. In this scenario, we focus on improving gross merchandise volume (GMV) and daily active buyers (DAB). The A/B experiment is conducted with a traffic allocation of 10\% vs. 10\% for 7 days. The results show that RedMall increases GMV by \textbf{1.08\%} and DAB by \textbf{0.95\%},
indicating that RedMall improves the effectiveness of e-commerce content distribution.

\begin{table}[h!]
\centering
\captionsetup{justification=centering}
\caption{Online A/B Experiments. The p-values for all metrics are below 0.05.}
\label{tab:redmall-online-metrics}
\begin{tabular}{lll}
\hline
Scenarios & Online Metrics & RedMall \\
\hline
\multirow{3}{*}{E-commerce Advertising.}
  & CTR  & +0.59\% \\
  & ADVV & +2.24\% \\
  & Cost & +0.56\% \\
\hline
\multirow{2}{*}{E-commerce}
  & GMV  & +1.08\% \\
  & DAB  & +0.95\% \\
\hline
\end{tabular}
\end{table}

\subsection{Ablation Study}
We conduct ablation studies on the industrial dataset to evaluate the individual modules proposed in RedMall. We use HitRate@50 as the evaluation metric. 
The training set contains 150 million users, each represented by a sequence of SIDs corresponding to items clicked during the preceding month. The evaluation set contains 16,000 users, with each user’s latest purchased item used as the target.

\begin{table}[h!]
\centering
\captionsetup{justification=centering}
\caption{Ablation results on the industrial dataset.}
\label{tab:ablation-industrial}
\begin{tabular}{c l c c}
\hline
 & Model & HitRate@50 & Improvement \\
\hline
Baseline & RedMall-NTP & 0.1416 & 0 \\
A        & RedMall-MTP & 0.1535 & +1.19\% \\
B        & A+MoE & 0.1668 & +2.52\% \\
C        & B+SID Encoder & 0.1832 & +4.16\% \\
D        & C+Dual Gated Attention & 0.1860 & +4.44\% \\
E        & D+Exposure-Aware Loss & 0.1948 & +5.32\% \\
F        & E+RL & 0.1997 & +5.81\% \\
\hline
\end{tabular}
\end{table}

\textbf{Effectiveness of Multi-Target Prediction}.
As shown in Table~\ref{tab:ablation-industrial}, replacing next-token prediction (NTP) with multi-target prediction (MTP) improves HitRate@50 from 0.1416 to 0.1535, corresponding to a relative gain of 1.19\%. 
This result suggests that supervising multiple target items under the same user context provides a more informative training signal than predicting only a single subsequent item. 
By exposing the model to several relevant items from each user's interaction sequence, MTP may help it capture a broader range of user interests and better support multi-item retrieval. The improvement provides empirical evidence for using multi-target supervision in RedMall.

\textbf{Effectiveness of Mixture of Experts}.
As depicted in Table~\ref{tab:ablation-industrial}, adding the sparse MoE architecture to RedMall-MTP improves HitRate@50 from 0.1535 to 0.1668, a relative gain of 2.52\%. This indicates that increasing the decoder's parameter capacity can improve its ability to model user--item relationships and generate relevant recommendations. 
The result is achieved with only two of the 12 experts activated during inference, suggesting that sparse expert activation can provide a stronger model while limiting the active computation. Together with the improvement from NTP to MTP, this result supports the complementary contributions of richer target supervision and increased decoder capacity.

\textbf{Effectiveness of SID Encoder}.
SID Encoder increases HitRate@50 from 0.1668 to 0.1832, a relative gain of 4.16\%. The SID Encoder explicitly fuses embeddings across SID layers and maps the fused representation back to layer-specific features, 
which are added to the corresponding encoder inputs. This provides the encoder with additional information about the hierarchical dependencies among SID layers, complementing the decoder's autoregressive token prediction. The observed improvement suggests that explicitly modeling cross-layer SID relationships helps RedMall retrieve users' purchased items more effectively.

\textbf{Effectiveness of Dual Gated Attention}.
Adding Dual Gated Attention to the model improves HitRate@50 from 0.1832 to 0.1860, a relative gain of 4.44\% over the baseline. This result indicates that gated attention and bidirectional modeling of the observed history contribute additional benefits beyond the preceding architectural components. 
The gates allow the decoder to adaptively modulate attention outputs, while bidirectional attention lets historical items incorporate information from both earlier and later interactions. Together, these mechanisms may help the model form more informative representations of user interests, consistent with the observed improvement in retrieval performance.

\textbf{Effectiveness of Exposure-Aware Loss}.
As shown in Table~\ref{tab:ablation-industrial}, Exposure-Aware Loss improves HitRate@50 from 0.1860 to 0.1948. This corresponds to a 4.73\% relative improvement over the preceding configuration and a 5.32\% gain over the baseline. 
By treating exposed but unclicked items as negative feedback, the loss supplements click-based positive supervision and helps the model distinguish clicked items from exposed alternatives that users did not select. The improvement suggests that incorporating non-click responses provides a useful training signal for reducing the likelihood of generating less-preferred items.

\textbf{Effectiveness of Reinforcement Learning}.
GRPO with the LambdaRank loss achieves a HitRate@50 of 0.1997. This is an increase of 0.49\% over the exposure-aware model (0.1948), corresponding to a 5.81\% gain over the baseline. The result suggests that RL provides additional benefit beyond supervised learning with exposure-aware feedback. 

\textbf{Effectiveness of Intra-SID Item Ranking}.
We conduct an online A/B test to evaluate the effectiveness
of the proposed intra-SID item ranking strategy.
In the ablated variant, all items associated with each
generated SID are retrieved sequentially until the
retrieval quota is exhausted, without applying
posterior-based item ranking.
The results show that removing intra-SID item ranking
reduces the ADVV improvement to only 1.1\%,
which is not statistically significant ($p > 0.05$).

Further analysis reveals that the downstream pass-through
rate without intra-SID item ranking is approximately
half of that achieved with the proposed strategy.
This degradation is primarily attributed to duplicate
filtering, which removes a substantial proportion of
retrieved items.
These findings suggest that posterior-based item selection
not only improves the quality and diversity of retrieved
candidates but also reduces redundant retrieval,
thereby improving the consistency between the retrieval
stage and the downstream recommendation pipeline.

\section{Conclusion}

In this paper, we presented RedMall, a generative recommender designed for large-scale e-commerce. RedMall co-designs SID tokenization, SID encoding, architecture, and model training. SID tokenization provides accurate e-commerce representations and end-to-end SID generation. The SID encoder explicitly models dependencies across SID layers, while MoE, dual gated attention, and multi-target prediction scale model capacity, strengthen sequence modeling, 
and support session-wise recommendation, respectively. We further introduced the Exposure-Aware Loss to incorporate exposed but unclicked items as negative feedback, and combined GRPO with a LambdaRank loss to align user preferences and improve the ranking of positive items. 
Large-scale online A/B tests show that RedMall delivers substantial gains in both e-commerce recommendation and e-commerce advertising. These results demonstrate the potential of generative modeling for large-scale e-commerce. In future work, we plan to explore how to integrate RedMall with the reasoning capabilities of large language models to support accurate user-intent reasoning.

\clearpage

\addcontentsline{toc}{section}{References}
\bibliography{redmall}
\bibliographystyle{plain}

\end{document}